\documentclass[onecolumn]{aastex631}

\usepackage[]{xcolor}
\shorttitle{Large Scale Helical Magnetic Field in M87}
\shortauthors{Pasetto, A. et al.}
\graphicspath{{./}{figures/}}

\begin{document}

\title{Reading M87's DNA: A Double Helix revealing a large scale Helical Magnetic Field}

\author{Alice Pasetto}
\affiliation{Instituto de Radioastronom\'{\i}a y Astrof\'{\i}sica (IRyA-UNAM), 3-72 (Xangari), 8701, Morelia, Mexico}
\email{a.pasetto@irya.unam.mx}

\author{Carlos Carrasco-Gonz\'alez}
\affiliation{Instituto de Radioastronom\'{\i}a y Astrof\'{\i}sica (IRyA-UNAM), 3-72 (Xangari), 8701, Morelia, Mexico}

\author{Jos\'e L. G\'omez}
\affiliation{Instituto de Astrof\'isica de Andaluc\'ia (IAA-CSIC), Glorieta de la Astronom\'ia s/n, E-18008 Granada, Spain}

\author{Jos\'e-Maria Mart\'{\i}}
\affiliation{Departament d'Astronomia i Astrof\'{\i}sica, Universitat de Val\`encia, E-46100 Burjassot (Val\`encia), Spain} 
\affiliation{Observatori Astron\`omic, Universitat de Val\`encia, E-46980 Paterna (Val\`encia), Spain} 

\author{Manel Perucho}
\affiliation{Departament d'Astronomia i Astrof\'{\i}sica, Universitat de Val\`encia, E-46100 Burjassot (Val\`encia), Spain} 
\affiliation{Observatori Astron\`omic, Universitat de Val\`encia, E-46980 Paterna (Val\`encia), Spain} 

\author{Shane P. O’Sullivan}
\affiliation{School of Physical Sciences and Centre for Astrophysics \& Relativity, Dublin City University, Glasnevin, D09 W6Y4, Ireland.}

\author{Craig Anderson}
\affiliation{Jansky fellow of the National Radio Astronomy Observatory, 1003 Lopezville Rd, Socorro, NM 87801 USA}
\affiliation{CSIRO Astronomy and Space Science, PO Box 1130, Bentley WA 6102, Australia}
\affiliation{ATNF, CSIRO Astronomy and Space Science, PO Box 76, Epping, New South Wales 1710, Australia}

\author{Daniel Jacobo D\'iaz-Gonz\'alez }
\affiliation{Instituto de Radioastronom\'{\i}a y Astrof\'{\i}sica (IRyA-UNAM), 3-72 (Xangari), 8701, Morelia, Mexico}

\author{Antonio Fuentes}
\affiliation{Instituto de Astrof\'isica de Andaluc\'ia (IAA-CSIC), Glorieta de la Astronom\'ia s/n, E-18008 Granada, Spain}

\author{John Wardle}
\affiliation{Physics Department, Brandeis University, Waltham, MA 02453, USA}


%



\begin{abstract}
We present unprecedented high fidelity radio images of the M87 jet. We analyzed Jansky Very Large Array (VLA) broadband, full polarization, radio data from 4 to 18 GHz. The observations were taken with the most extended configuration (A configuration), which allow the study of the emission of the jet up to kpc scales with a linear resolution $\sim$10 pc. The high sensitivity and resolution of our data allow to resolve the jet width. We confirm a double-helix morphology of the jet material between $\sim$300 pc and $\sim$1 kpc. We found a gradient of the polarization degree with a minimum at the projected axis and maxima at the jet edges, and a gradient in the Faraday depth with opposite signs at the jet edges. We also found that the behavior of the polarization properties along the wide range of frequencies is consistent with internal Faraday depolarization. All these characteristics strongly support the presence of a helical magnetic field in the M87 jet up to 1 kpc from the central black hole although the jet is most likely particle dominated at these large scales. Therefore, we propose a plausible scenario in which the helical configuration of the magnetic field has been maintained to large scales thanks to the presence of Kelvin-Helmholtz instabilities. 
\end{abstract}



\section{Introduction} 

Active Galactic Nuclei (AGN) are powerful astrophysical objects, emitting at all wavelengths. One of the most intriguing features observed in AGNs are the powerful relativistic plasma jets which emit synchrotron radiation \citep[for the description of the original phenomenological model of AGN jets see ][]{Rees1971,Scheuer1974}. Several models have been proposed to explain how AGN jets are launched and collimated, and there is a general consensus that a strong magnetic field plays a fundamental role \citep[see recent review by ][and references therein]{Blandford2019}. The most accepted mechanisms assume magneto-centrifugal launching near the black hole (BH), either from a rotating accretion disk or from the BH ergosphere, and subsequent collimation of the material into a narrow jet due to confinement of the material by a helical magnetic field \citep[see ][]{BlandfordPayne1982, BlandfordZnajek1977}. Obviously, understanding how AGN jets work requires of mapping the 3D configuration of the magnetic field in a wide range of scales spanning several orders of magnitude, from the event-horizon of the BH (event-horizon scales), through the accretion disk and vicinity of the BH ($\lesssim$1~pc), up to the typical lengths of jets ($\sim$1~kpc) \citep[for a recent review on magnetic field in AGN, see ][and references therein]{Pudritz2012, Gabuzda2018}. 

It is very convenient that the material in relativistic jets emits lot of energy through synchtrotron emission. This emission is intrinsically linearly polarized and the analysis of its properties allows to perform detailed studies of the magnetic field configuration in AGN jets: the total intensity is related to the strength of the magnetic field, while the polarization direction is related to the orientation of the magnetic field projected on the plane of the sky \citep[for details see, e.g., ][]{Wardle2001, Wardle2003}. Moreover, from the analysis of the spatial and spectral variations of the polarization properties, it is also possible to infer the 3D configuration of the magnetic field \citep[see for details:][]{Laing1981, Blandford1993, Lyutikov2005}. In particular, for a helical magnetic field, we expect gradients of the Faraday depth (which causes a rotation of the polarization with wavelength) and polarization degree \citep[fraction of polarized emission, see recent simulations by][]{Moya2021,Fuentes2021} across the width of the jet. Note that these spatial gradients of Faraday Depths could also be mimicked with the presence of external material to the jet which only rotates the polarization angle. Therefore, in order to be interpreted as a solid sign of a helical configuration, the spatial gradients of Faraday Depth should be detected at several places along the jet direction, and it would be desirable to rule out the presence of external material, and/or to identify that the polarization properties are tracing the internal magnetic field. However, this is not an easy task since it requires of very high angular resolution to resolve the jet width, very high sensitivity to detect polarized emission, and observations at several wavelengths to study polarization properties across the spectrum \citep[][]{Broderick2010, Taylor2010, Hovatta2012}. 

Currently, the highest angular resolutions which allow the study of the formation and collimation of AGN jets can only be reached by using the Very Long Baseline Interferometry (VLBI) technique \citep[for historical and technical details, see ][]{Thompson2017}. Very recently, observations performed with the Event Horizon Telescope (EHT) obtained polarized data on event-horizon scales, mapping, for the first time, the magnetic field close the supermassive BH at the center of M87 and favouring launching of the jet from the ergosphere of the BH \citep{EHTVIII2021}. Concerning the jet collimation, several works using VLBI facilities have obtained evidences of helical magnetic fields present within the first few parsecs in several bright objects \citep[e.g.,][]{Asada2002,Gabuzda2004,Gomez2008,Taylor2010,Hovatta2012,Mahmud2013,Gabuzda2015,Gabuzda2017}. However, the question of up to which scales is the helical magnetic field present and, more importantly, actively collimating the jet, is still open. In recent years, the Karl G. Jansky Very Large Array (VLA) has emerged as a powerful instrument to study the polarization properties across the spectrum thanks to its capability of observing wide ranges of wavelengths with very high sensitivity. This allows the use of sophisticated tools in order to infer properties of the magnetic field and the material of the jets \citep[e.g.][]{OSullivan2012,Anderson2016, Pasetto2018}. Although the VLA cannot reach the high angular resolution performed by VLBI interferometers, it is still able to resolve the structure at kpc scales of nearby objects and, therefore, to expand our knowledge on the structure of AGN jets to larger scales.

One of the best targets for the study of the magnetic field configuration is the jet emanating from M87. This is one of the nearest powerful relativistic jets \citep[16.7Mpc;][]{Mei2007} and has been intensively studied at several wavelengths and physical scales. The collimated jet has a total extension of $\sim1$ kpc and shows a succession of bright knots in optical and radio images \citep[see e.g.][]{Biretta1993, Perlman1999}. Early sensitive radio images revealed complex polarization structures in some of the knots of M87 and proposed the presence of twisted filament-like structures which seem to be consistent with a double-helix morphology \citep{Owen1980, Owen1989, Owen1990}. Several formation mechanisms have been discussed to explain these structures \citep[ e.g. the 'Sweeping Magnetic-Twist' model, ][]{Nakamura2001}, but the most likely explanation seems to be Kelvin-Helmholtz (K-H) instabilities in the jet \citep{Lobanov2003, Hardee2011}. Several efforts have been made at radio wavelengths to infer the 3D structure of the magnetic field at different spatial scales \citep[e.g.,][]{Chen2011, Avachat2016, Algaba2016, Park2019, Kravchenko2020}. However, although most of these works allow a study of the polarization properties across the radio spectrum, currently there are no reported observations with enough angular resolution to study these properties across the jet width. Thus, despite several theoretical works claiming its presence at different scales \citep[e.g., ][]{Nakamura2010, Nakamura2014}, we still lack a clear observational evidence of a helical magnetic field in the M87 jet.

Here, we present unprecedented high fidelity VLA radio images of the M87 jet which allow us to fully resolve the structure of the jet and to perform a detailed analysis of the polarized spectrum. In our images, the filamentary structure of the jet is clearly revealed as a double helix. Our observations allow to separate emission from the edges and the central axis of the jet, and thus to study the polarization properties across the jet width. The large bandwidth used in our data allows a detailed modeling of the depolarization effects in the jet. The main result of our analysis is that we can infer a helical configuration of the magnetic field present up to 1-kpc from the core of M87, thanks to the detection of Faraday depth gradients transversally to the jet flowing direction. This is the most clear evidence to date for a helical configuration of the magnetic field in an AGN jet at these scales.

\section{Observations and Data Analysis}
\label{DA}

\subsection{VLA Observations}

 Observations were performed with the VLA of the National Radio Astronomy Observatory (NRAO)\footnote{The National Radio Astronomy Observatory is a facility of the National Science Foundation operated under cooperative agreement by Associated Universities, Inc.} in a single session on August 8th, 2015 (NRAO Project Code: 15A-170) at bands C, X, and Ku using the A configuration. We used the standard continuum mode (2 MHz width channels) covering the full available frequency range at each band: 4-8 GHz at C band, 8-12 GHz at X band, and 12-18 GHz at Ku band. Thus, the combined data fully covers a total bandwidth of 14 GHz. We observed in a single long run where we alternated several times between the different bands. In this way, at each band, the observation covers a wide range of parallactic angles which allows calibration of the polarization leakeage. At each band, we observed $\sim$1~hour on-source. We observed alternately every $\sim$1 min, two pointings separated by 12.4$\arcsec$, one centered on M87's core (J2000 12$^h$30$^m$49.42$^s$ 12$^\circ$23$\arcmin$28.0$\arcsec$), the other centered on knot A (J2000 12$^h$30$^m$49.46$^s$ 12$^\circ$23$\arcmin$32.0$\arcsec$). We also observed 3C286 which was used to calibrate flux density, bandpass and polarization angle. The data were calibrated using the Common Astronomy Software Applications (CASA) package (version 5.1.2) and we used the latest available version at that moment of the NRAO pipeline for VLA continuum data which was modified to include polarization calibration after the complex gain calibration. Less than 1\% of the data at each band were flagged due to interferences and/or errors (0.6\% at C band, 0.5\% at X band and 0.9\% at Ku band).

 For the polarization calibration of the three wide bands, we followed a procedure similar to that used in \cite{Pasetto2018}. Essentially, we used a polarized model for 3C286 which takes into account variations of the stokes Q and U parameters within the whole frequency range observed \cite{Perley2013P}. Then, we obtained solutions for the polarization at each 2 MHz spectral channel. First, we correct for the cross-hand delays using 3C286. The leakeage terms were corrected using the core of M87, observed in a large range of parallactic angles, and assuming it to have an unknown polarization value. Finally, to set the absolute electric vector polarization angle (EVPA), the polarized model of 3C286 has been used. The pipeline was run for each band separately. Once calibrated, we concatenated all data into a single file to produce images and to self-calibrate. During self-calibration, models were obtained from uniform weighted images, which resolves out most of the very large extended emission, but at the same time allows us to detect the collimated jet with a high signal-to-noise ratio (S/N). We performed several cycles of imaging and self-calibration (amplitude and phase) until the quality (rms noise and SNR) of the resulting maps does not seem to change.

 All the images were produced by using the task \emph{tclean} from CASA as a 2-pointing mosaic. For all images using large bandwidths, we used multi-scale multi-frequency synthesis \citep[MS-MFS; see][]{Rau2011}, which takes into account possible changes in the flux density with frequency. The spectrum of each flux component is modeled by a Taylor expansion about the reference frequency, $\nu_0$=11~GHz. We chose the parameter nterms=2 in \emph{tclean}, which results in an expansion of first order. This is equivalent to assuming a spectral power-law variation of the flux ($S_\nu$) with frequency ($\nu$) in the form $S_\nu \propto \nu^\alpha$, where $\alpha$ is an average spectral within the used bandwidth. This is the expected form of the spectrum in the observed frequency range, i.e., optically thin synchrotron emission. From the self-calibrated data-set, we made final images using the total 14 GHz bandwidth and different weightings of the uv plane, robust=1 and uniform. The robust=1 weighted image has a final resolution of $\sim$0$\farcs$2, while the uniform weighted image $\sim$0$\farcs$09. For the modeling of the polarization properties in wavelength (see also \S\ref{DepolModel}), we made several uniform weighted images of Stokes parameters I, Q, and U in smaller 128 MHz wide sub-bands. Thus, we produced 112 sub-bands in order to cover the total 14 GHz frequency range. Since the M87 jet shows emission at different scales, all sub-band images were restricted to the same uv range, from 20 to 600 k$\lambda$. This results in sub-band images sensitive to the same spatial scales and with a similar resolution. Nevertheless, for the final analysis, we convolved all these sub-band images to a circular beam of 0$\farcs$438 to ensure proper comparison. 

\subsection{Depolarization modeling} \label{DepolModel}
 
The synchrotron radiation from a radio source is partially linearly polarized and its polarization signal can be represented by the Stokes parameters, I, Q, and U, which can be combined to describe the polarized signal as a complex number \citep[see e.g.,][]{Sokoloff98}:

\begin{equation}
P=Q+iU=pIe^{2i\chi}
\end{equation}

\noindent where $p=P/I$ is the fractional linear polarization, i.e., the amount of emission linearly polarized, and $\chi$ is the polarization angle, which is perpendicular to the direction of the component of the magnetic field projected to the plane of the sky. When the polarized radiation passes through a magneto-ionic medium, it suffers a rotation of the polarized plane of the linearly polarized wave, an effect known as Faraday rotation. The amount of rotation depends on the wavelength and the physical parameters of the medium through which the polarized light has passed. The latter are included in an useful parameter known as Faraday depth:

\begin{equation} \label{FarDepth}
\phi [rad/m^{-2}] = 8.1 \times 10^5 \int n_e \ B_{los} dL
\end{equation}

\noindent where $n_{e}$ [cm$^{-3}$] is the  electron density of the thermal electrons in the plasma, $B_{los}$ [Gauss] is the magnetic field component along the line of sight and $L$ is the path length [pc]. The amount of rotation is known to be wavelength-dependent. Classically, Faraday rotation has been described as a simple linear variation of the polarization angle with the square of the observed wavelength,

\begin{equation}
\chi_{obs} = \chi_0 + \phi \lambda^2,
\end{equation}

\noindent or, using the complex notation,

\begin{equation}
P(\lambda)=p_0e^{2i (\chi_0+ \phi \lambda^2)},
\label{FaradayScreen}
\end{equation}

\noindent where $\chi_{obs}$ is the observed polarization angle, and $\chi_0$ is the intrinsic polarization angle at which the light was emitted from the synchrotron source. Note that this equation implies only a rotation of the polarization angle, but not a change in the total polarized emission, and thus, the fractional polarization remains constant over the whole spectrum. However, this behavior is only true in a very simple case in which the polarized emission from a synchrotron source passes through a plasma, physically disconnected from the synchrotron source, and containing uniform magnetic field and electron density. For this reason, the Faraday depth ($\phi$) in equation \ref{FaradayScreen} is usually replaced by the Rotation Measure (RM), which is actually the emission-weighted mean value of the Faraday depth, i.e., $RM = <n_e> <B_{los}> L$ in which positive and negative RM signs are considered. However, real cases are much more complicated and result in complex variations of the polarization angle and the fractional polarization. In the literature, there are several theoretical descriptions of more realistic cases which consider the rotating material to be either external or internal to the synchrotron emitter, and the magnetic field ordered and/or turbulent \citep[see recent review by][]{Pasetto2021}. Since all these scenarios predict a decrease of the fractional polarization at longer wavelengths as a consequence of the Faraday rotation, their effect in the polarized spectrum is known as Faraday depolarization. It is only since very recently, when powerful radio interferometers have been upgraded with wideband receivers, that it has been possible to observe and model complex behaviors of the fractional polarization and the polarization angle in synchrotron sources \citep[e.g., ][]{OSullivan2012, Anderson2016, Pasetto2018}. 


Depolarization effects can still be important in the case of well-resolved relativistic jets, particularly in the presence of internal Faraday rotation. Particles in the jet emit linearly polarized radiation perpendicular to the local magnetic field. However, depending on the orientation of the jet relative to the observer, one should expect different results for light coming from different regions in the jet. For example, for a cylindrical or conical jet with its axis oriented parallel to the plane of the sky, the emission that travels through the jet axis will suffer stronger internal depolarization due to the longer path length through the jet (both due to internal Faraday rotation and the changing field orientation through the jet). Beam depolarization can also occur due to limited transverse resolution. In this work, we take advantage of the availability of high sensitivity, high angular resolution observations of a resolved jet covering a wide range of wavelengths to perform detailed modeling of the polarization properties. For this, we model the behavior of the polarization properties at each pixel using the technique known as QU-fitting \citep[][]{OSullivan2012, Pasetto2018}. We used a model of internal depolarization which is adequate to describe the emission from a jet, i.e., a magnetized plasma which is emitting and rotating the polarization angle. This model (Slab model) is described in \citep{Burn1966} and predicts that the polarization emission follows the equation:

\begin{equation}
P(\lambda)=p_0e^{2i(\chi_0+\frac{1}{2}\phi\lambda^2)}\left(\frac{{\rm sin} \phi\lambda^2}{\phi\lambda^2}\right),
\label{Burn66}
\end{equation}
\noindent
where the $\phi$ is the Faraday depth through the region. The model describes the scenario in which, the emitting and rotating regions are co-spatial in the presence of a regular magnetic field. In this case, the radiation coming from the most distant part of the region (with respect to the observer) undergoes a different amount of Faraday rotation with respect to the radiation coming from the nearest part of that region.
This scenario, although still simple, describes the polarization properties behavior in $\lambda^2$ more properly than a single simple Faraday screen (, i.e. equation \ref{FaradayScreen}), which does not account for Faraday depolarization effects. Note that in a scenario in which a large amount of ionized gas is in front of the emitting material, we could also expect external depolarization. However, this would produce a constant RM across the jet and, most likely, strong depolarization \citep[see e.g.][]{Knuettel2019}. Fortunately, in this case the surrounding gas and dust has been mapped \citep[see][]{Sparks1993}. A large and massive cloud has been found associated to M87. This structure seems to be in foreground only at the position of the core \citep[see also][]{EHTVII2021,EHTVIII2021}, while the rest of the cloud is not intersecting the jet. According to this result and given the high resolution and high frequency of this study, it is safe to assume that the polarization behaviour across the spectrum is dominated by internal depolarization.

As described in the previous section, we made 112 images of sub-bands at different wavelengths. Thus, at each pixel, we modeled the behavior of $Q(\lambda^2)$ and $U(\lambda^2)$. First, we used equation \ref{FaradayScreen} which describes the simplest scenario, a simple Faraday screen. The parameters, \textit{$p_0$}, \textit{$\chi_0$} and \textit{RM}, obtained from this first modeling, were then used as initial condition for a second modeling using the internal Faraday depolarization model in equation \ref{Burn66}. From this procedure, we obtained resolved maps of the three polarization parameters: $p_0$, $\chi_0$ and $\phi$ for each model. We note that both models resulted in very similar maps of all these parameters. However, the simple Faraday rotation screen is not able to represent the observed depolarization at long wavelengths in several pixels with high signal-to-noise ratios (see \S\ref{RD}). Moreover, the high frequency and high angular resolution of these observations restrict the emission to the most compact synchrotron components, with a linear scale of $\sim$10-60 pc. Moreover, as commented above, the absence of intervening material in the foreground of the jet, supports the idea of internal Faraday depolarization. This is also supported by the complex behaviour of the polarized spectrum (see \S\ref{RD}) From the intrinsic polarization angle (i.e., the polarization angle corrected by Faraday rotation), we obtained a map of the magnetic field component parallel to the plane of the sky by simply rotating 90$^\circ$. The magnetic field vector map is then transformed into a streamline image by using the Line Integral Convolution (LIC) technique \citep[][]{Cabral1993}. We also performed a depolarization model fit in two defined areas well separated from each other (see \S\ref{RD}).

\section{Results and discussion}
\label{RD}

\begin{figure*}[ht]
\begin{center}
\includegraphics[width=1.\textwidth]{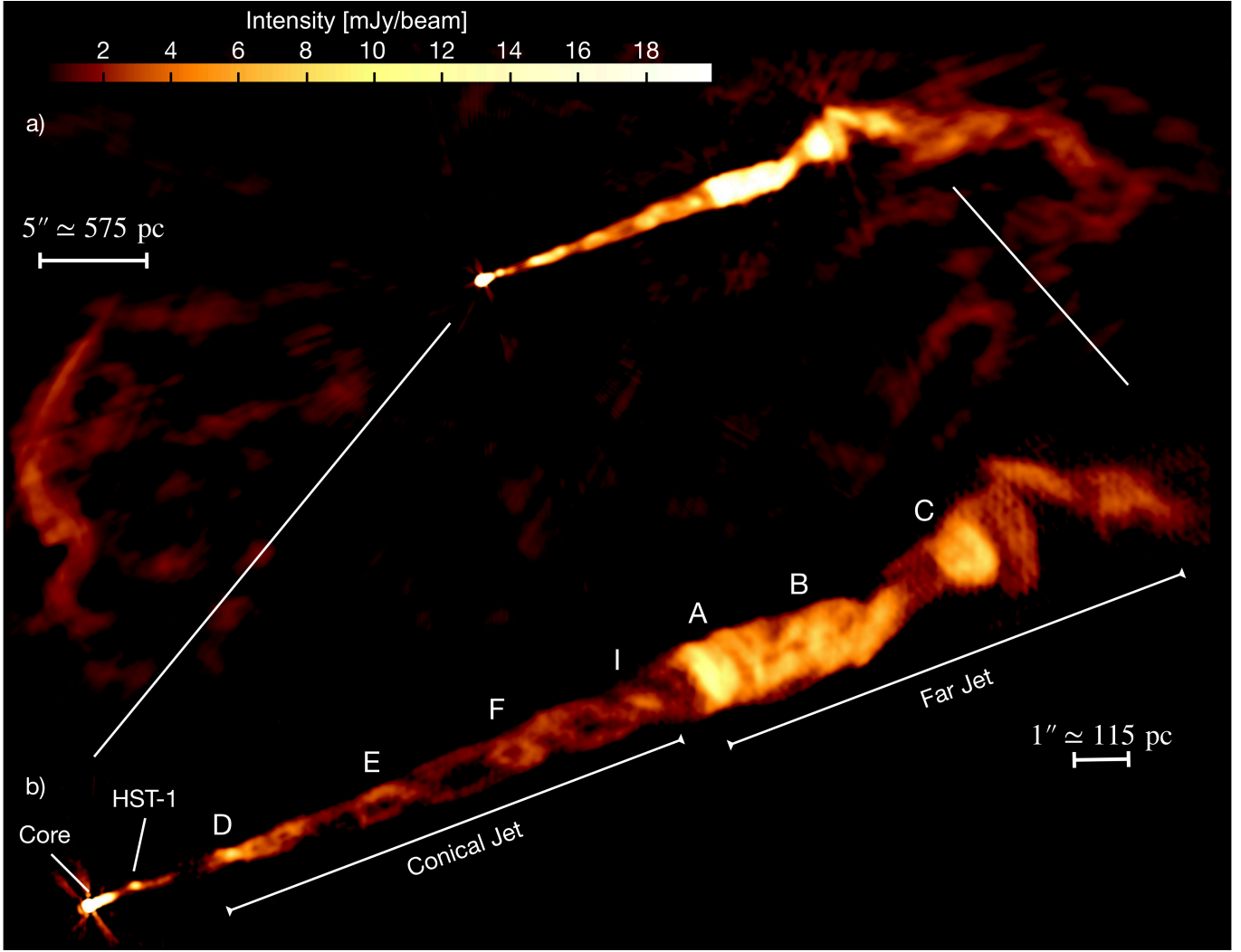}
\caption{The M87 VLA radio jet combining all the available frequencies (from 4 to 18 GHz). a) Image with an angular resolution of 0$\farcs$2 (robust=1) sensitive to large-scale emission. It shows the well-known morphology consisting of a core, a highly collimated conical jet ending in a series of bright knots, and large extended lobes. b) Image with a higher angular resolution of 0$\farcs$09 (uniform weighted), and it shows the structure of the collimated jet in great detail. Several knots, previously identified in optical images, are labeled. This image shows a clear double-helix structure in the conical jet.}
\label{Fig1}
\end{center}
\end{figure*}

Our new VLA observations offer a view of the M87 jet with an unprecedent combination of high sensitivity and high angular resolution (see Figure \ref{Fig1}). Not only we detect emission from the bright collimated jet, but we are also able to detect very extended low brightness emission from the terminal lobes (see Figure \ref{Fig1}a). Our highest angular resolution (0$\farcs$09) image allow us to focus on the $\sim$1~kpc structure of the collimated jet at a lineal resolution of $\sim$10~pc (see Figure \ref{Fig1}b). The jet arises highly collimated from the core of M87 up to the radio knot HST-1. From this point, the jet width seems to increase as a conical jet (see Figure \ref{Fig1}b) until it reaches the bright radio knot A at $\sim$1~kpc from the core. The conical part of the jet includes several knots previously identified from optical and radio images (knots D, E, F, and I). Our high angular resolution allows us to resolve filaments of material that are waving together as ribbon-like, all along the conical jet (see Figure \ref{Fig2}a). Substructures in this part of the jet were previously studied from lower sensitivity images which pointed out oscillations in the brightness of the filaments \citep[][]{Owen1989, Biretta1993}. Note that from our image (see Figure \ref{Fig1}b), it is clear that the previously named knots E, F and I are actually regions in the filaments which becomes brighter. It has been proposed that this filamentary morphology is the result of Kelvin-Helmholtz (K-H) instabilities in the jet \citep{Bicknell1996, Lobanov2003, Hardee2011}. Moreover, the filaments seen in our image confirm the double-helix structure extending from HST-1 to knot A proposed by \cite{Lobanov2003} and \cite{Hardee2011}. The use of K-H linear modes to estimate the physical conditions in the jets have been used with success in a number of cases besides the jet of M87, including the twisted structures in the 3C~120 jet \citep{Hardee2005} and, remarkably, in the 3C~273 jet \citep{Lobanov2001, Perucho2006}. The analysis of the K-H instability has also been used to interpret the transversal structure in the 0836+710 jet \citep{Perucho2007} as well as its helical structure \citep{Perucho2012,Vega2019}.

\subsection{A large scale helical magnetic field revealed}

The presence of a helical filamentary structure in the jet does not imply the presence of a helical magnetic field in the jet. In order to infer the 3D configuration of the magnetic field, we need to spatially and spectroscopically study the polarization properties of the emission. It is well known that, for a helical magnetic field, it is expected to detect gradients across the jet width of both, the fractional polarization and the Faraday depth \citep[][]{Lyutikov2005, Gomez2008}. The emission coming from the projected jet axis is expected to suffer from strong depolarization as the light travels across the jet width with a changing direction of the magnetic field lines and electron density. Thus, we expect to detect a minimum of the fractional polarization at the center of the jet axis \citep[scenario also supported by recent simulations, see ][]{Moya2021,Fuentes2021}. In contrast, emission from the projected edges of the jet, will not suffer from strong depolarization since their emission is passing through smaller regions where the magnetic field and electron density can be consider constant. Moreover, a helical magnetic field implies the presence of a toroidal component of the magnetic field, i.e., at the edges of the jet, the line-of-sight component of the magnetic field should show opposite directions. Since the Faraday depth depends on the component of the magnetic field parallel to the line of sight (see equation \ref{FarDepth}), a toroidal magnetic field could be identified as a gradient of the Faraday depth across the width of the jet. From these considerations, it is clear that in order to discern the 3D configuration of the magnetic field, it is essential to resolve polarization across the jet width. However, one common problem is that the polarization study should be performed at lower angular resolution than the stokes I image because, in order to compare polarization images at several wavelengths, we need to convolve all of them to the lower resolution image, which is defined by the longer wavelength image. In our case, although the stokes I image has a resolution of 0$\farcs$09, the polarization study is limited to a resolution four times lower, $\sim$0$\farcs$4, which allows to resolve the jet width with 3 beams. However, in this particular case, thanks to the presence of the double-helix morphology, we can better study the polarization properties across the jet width. As we move along the jet direction, regions where the emission is dominated by the edges of the jet take turns with regions where most of the emission is coming mainly from the projected jet axis. This allows us to better resolve polarization properties across the jet width as we move along the jet direction. 

\begin{figure*}
\begin{center}
\includegraphics[width=0.7\textwidth]{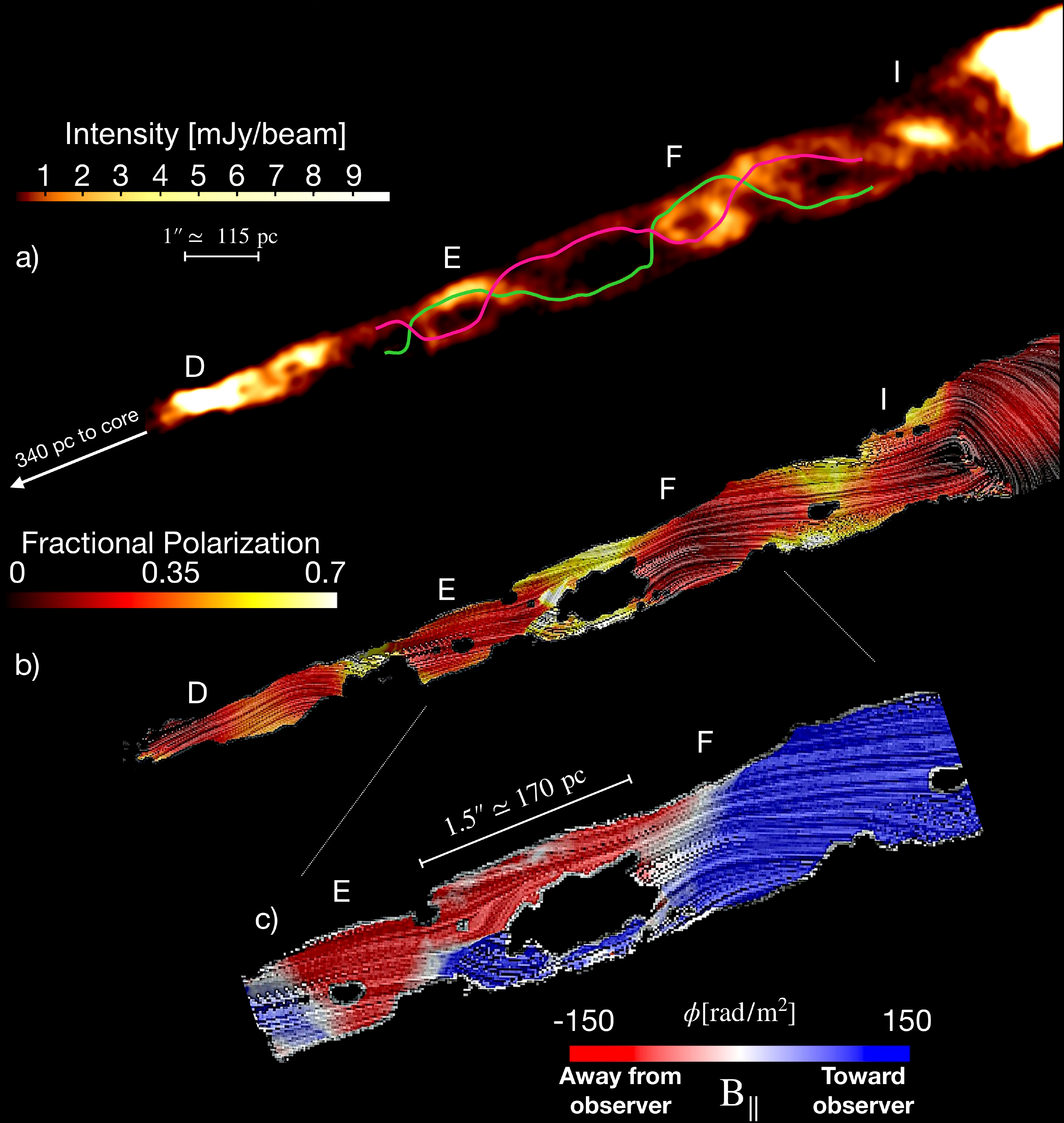}
\caption{Analysis of the polarization properties in the conical jet of M87 revealing a helical magnetic field configuration. The central frequency is  11 GHz. At the top, there is the high angular resolution (0$\farcs$09) Stokes I image showing the double-helix structure between knots D and I. The two lines were obtained by fitting slices perpendicular to the jet axis to two Gaussians. The middle image shows the LIC streamline image of the component of magnetic field parallel to the plane of the sky over the fractional polarization map. The magnetic field lines approximately follows the double-helix structure which suggests we are observing the poloidal component of the magnetic field in the jet. Note the increase of the fractional polarization when emission from the edges of the jet width can be well resolved. The bottom image shows a close-up to the region enclosing knots E and F, the region where the filaments appear more separated. Again, we show the streamline image of the magnetic field parallel to the plane of the sky, but now, the color scale shows the values of the Faraday depth obtained from our modeling of the Stokes parameters $Q(\lambda^2)$ and $U(\lambda^2)$. The sign of the Faraday depth traces the direction of the magnetic field along the line of sight. Thus, it can be clearly seen that in the region where we are able to separate emission from both edges of the jet, magnetic field has opposite directions, strongly suggesting a toroidal component. All these characteristics are strongly supporting the presence of a helical configuration in the M87 jet.}
\label{Fig2}
\end{center}
\end{figure*}

\begin{figure*}
\begin{center}
\includegraphics[width=0.7\textwidth]{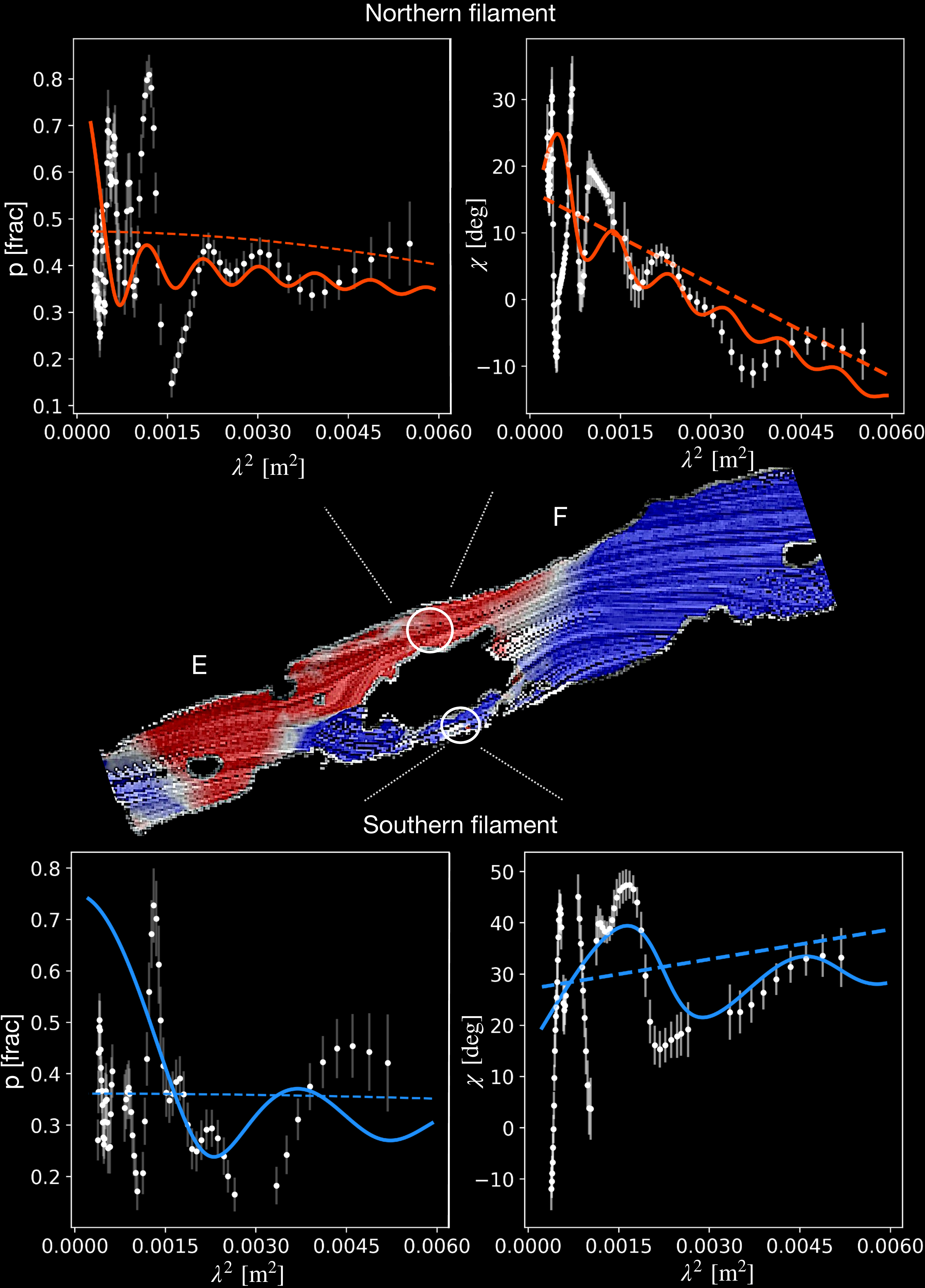}
\caption{Panels show the behavior of the fractional polarization and polarization angle in $\lambda^2$ in the two regions, marked by the two white circles in the RM map, located at the filaments. At the resolution of our observations, it is expected that the polarization behaviour is dominated by internal effects in the material of the jet. Thus, we first used a simple internal Faraday screen model, equation \ref{Burn66}, which is represented by the dashed lines and which is able to describe the global behaviour of the polarization angle. Note, however, that the data show more complex behaviour at smaller scales. Remarkably, strong depolarization is observed across the spectrum, and there are oscillations of the polarization angle at several wavelengths. These characteristics strongly support the presence of internal depolarization effects. However, a full description of the polarized spectrum requires of more complicated physically based models. Therefore, the solid lines show a scenario assuming two internal Faraday screens which better describe the behavior of the data. However, we emphasize that these are not fits to the data, and that more specific, physically based models are necessary to explain the spectral changes in polarization parameters.}
\label{Fig3}
\end{center}
\end{figure*}

In Figures \ref{Fig2}b and \ref{Fig2}c, we show a superposition of the LIC map of the magnetic field projected to the plane of the sky over the fractional polarization and over the Faraday depth obtained from our depolarization modeling. First, note that where the filaments are well separated (e.g., between knots E and F and between knots F and I) the magnetic field is well ordered reaching high fractional polarization values of $\simeq$0.5. Larger values, almost $\simeq$0.7, are detected where the magnetic field lines open like a funnel (between knots E and F) following the two filaments. On the contrary, where the filaments intersect each other the emission suffers for strong depolarization effects and the fractional polarization drops to lower values of less than $\simeq$ 0.1 (e.g., at the position of knot F and knot I). Second, note that in the region between knots E and F, where the filaments are well separated, we detect a Faraday depth gradient from negative (red, B-line moving away from the observer) to positive (blue, B-line moving toward the observer), evidencing a toroidal component of the magnetic field (see Figure \ref{Fig2}c). We also analyzed in more detail the polarized emission at two regions well separated in the filaments between knots E and F (named the Northern and Southern filament, respectively). This is shown in Figure \ref{Fig3}, where we show the values of the fractional polarization and the polarization angle as a function of $\lambda^2$ and two depolarization models. Dashed lines show the result of the modeling considering a single internal Faraday screen. This is the model used to generate the images of fractional polarization, Faraday depth and the magnetic field (from the polarization angle map). As can be seen, although this model is able to represent the global variation of the polarization parameters in the observed range of wavelengths, their detailed spectral behavior is actually much more complex. In particular, we note strong depolarization effects and complex changes in the polarization angle across the spectrum (see Figure \ref{Fig3}). These effects strongly support the presences of internal Faraday depolarization in the jet, and a proper modeling of the polarized spectrum requires of a scenario more complex than a simple internal Faraday screen. In Figure \ref{Fig3}, we show that a slightly more complex scenario consisting of two internal Faraday screens could better represent the observed fine structure in the spectrum. This ad-hoc scenario could be the consequence of changes in the physical conditions along the line-of-sight or within the beam size. However, we emphasize that these models are not a fit, but only an exercise to show that a more complex scenario involving internal Faraday screen is necessary. Actually, a more realistic, physically based model of the jet emission should be performed.

\begin{figure*}
\begin{center}
\includegraphics[width=0.8\textwidth]{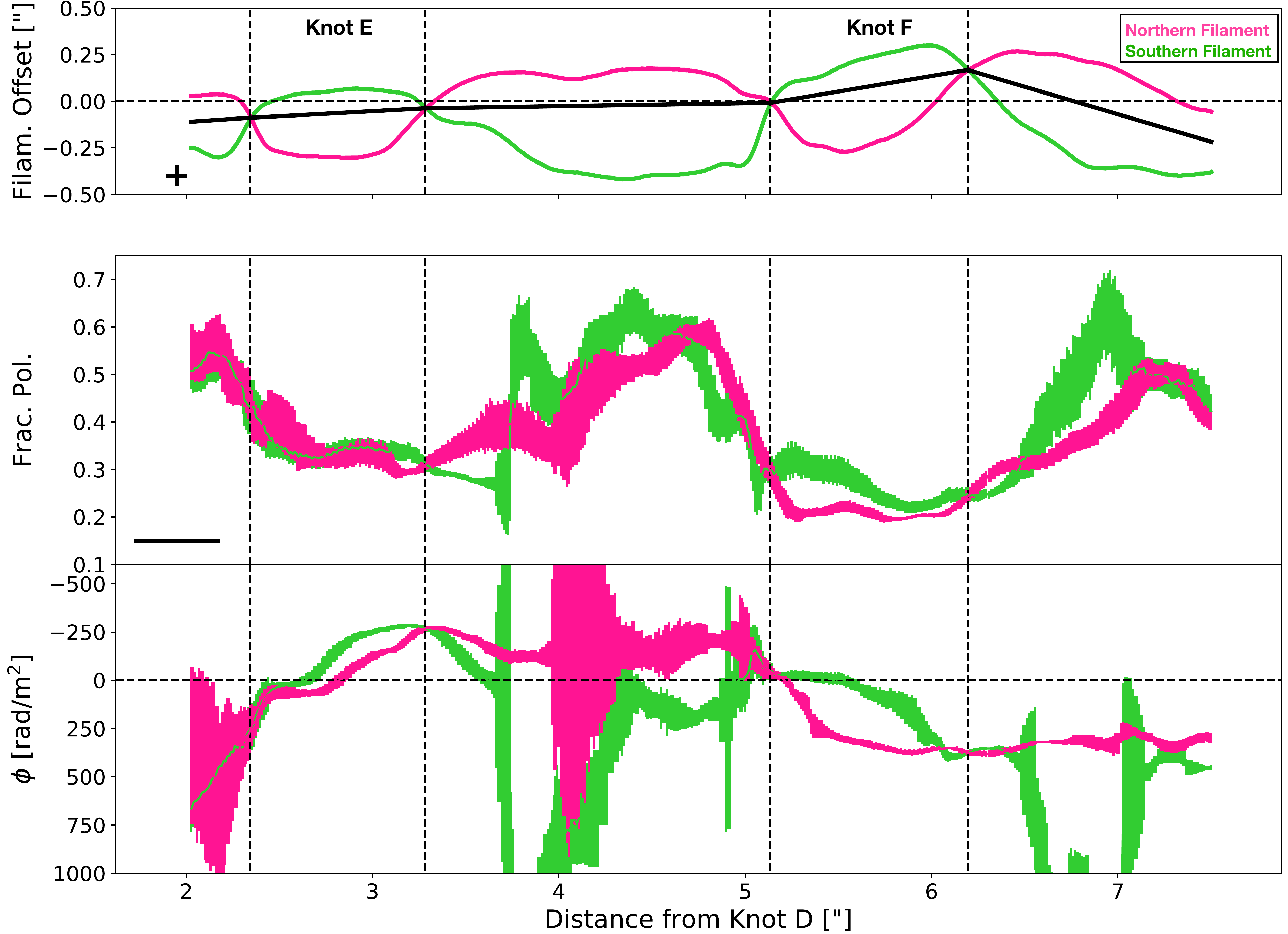}
\caption{Analysis of the polarization properties along the filaments. The top panel shows the positions of the filaments with respect to the observed jet axis (defined as the line joining knots D and A). The black solid line is obtained by joining the points where both filaments cross, and suggests that the real jet axis shows some bending. The middle panel shows the values of the fractional polarization averaged over a region of 5 pixels around the central position of the filaments. The width of the lines is the standard deviation. In a similar way, the bottom panel shows the averaged Faraday depth along the filaments. Note that the values in the y-axis are inverted; we show values from positive at the bottom to negative at the top. This has been done only to follow the same sign of Faraday depth shown in the previous RM map.}
\label{Fig4}
\end{center}
\end{figure*}

Finally, in Figure \ref{Fig4}, we present an analysis of the fractional polarization and the Faraday depth (resulted from our single internal Faraday depolarization model fit) along the two filaments. Note that we found gradients of the Faraday depth also where the filaments are separated at the position of knot E and knot F. However, these gradients do not show a change in sign between the two filaments; Faraday depth values are all negative in knot E and all positive in knot F. This is still consistent with a helical magnetic field in the case that the jet axis at these positions is not parallel to the plane of the sky \citep{Pudritz2012}. Indeed, we found that the filaments are not fully aligned with the global jet direction (see dashed line in Figure \ref{Fig4}-top panel), suggesting that some bending or even twisting is occurring in the jet axis itself following the instabilities, specially when approaching to knot A. Interestingly, the filaments between knots E and F, where a gradient from positive to negative is found, seem to be well aligned with the jet axis. Therefore, also the gradients in Faraday depth detected at knot E and knot F are most likely tracing the presence of a toroidal component of the magnetic field. In summary, we found that both, the spatial and spectral polarization properties are fully consistent with the expectations of a helical magnetic field present in the M87 jet between $\sim$300~pc and $\sim$1 kpc.

\subsection{Is the magnetic field actively collimating the jet up to 1~kpc?}

The striking alignment of the magnetic field lines with the filaments observed between knots D and I could be interpreted as an indication of the magnetically dominated character of the M87 jet at kpc scales. This is the model proposed by \cite{Nakamura2001}, \cite{Nakamura2004} to explain the production of wiggled structures in AGN jets as those observed in the M87 jet. This would imply that the M87 jet is being magnetically dominated up to $\sim$1~kpc from the core. However, this interpretation challenges the present paradigm for the formation and acceleration of jets, which implies a gradual change in the jet flow from magnetically dominated to kinetically dominated as the acceleration progresses and the electromagnetic energy is transferred to the plasma \citep{Vlahakis2003, Vlahakis2004, Beskin2006, Lyubarsky2009}. In the following, we explore this possibility.

From VLBI observations, an upper limit to the magnetic field of $B_{\rm VLBI}\simeq$0.2~G has been obtained for the core of M87 with a size of $R_{VLBI}\simeq$0.03~pc \citep[][]{Reynolds1996}. Then, flux conservation arguments allow us to estimate the magnetic field of the M87 jet at kiloparsec scales as:

\begin{equation}
\label{eq:bkpc}
B_{\rm kpc} \approx \left(\frac{\Gamma_{\rm VLBI}}{\Gamma_{\rm kpc}}\right)^\mu \left(\frac{R_{\rm VLBI}}{R_{\rm kpc}}\right)^\nu B_{\rm VLBI},
\end{equation}

\noindent with $(\mu, \nu) = (1,1)$ for a magnetic field predominantly toroidal, and $(\mu, \nu) = (0,2)$ for a magnetic field predominantly poloidal. In the previous expression, $R_{\rm kpc}$ is the jet radius at kiloparsec scales ($\sim 30$ pc, between HST-1 and knot A), and $\Gamma_{\rm VLBI}/\Gamma_{\rm kpc}$ is the ratio between the jet flow Lorentz factors at VLBI and kiloparsec scales. Analysis of the kinematics of the M87 jet shows that plasma in the jet accelerates from subluminal to superluminal speeds mostly before HST-1 (at $\sim 70$ pc from the core) \citep[see, e.g.,][]{Park2019} and then decelerates (due probably to mass entrainment) to subluminal (subrelativistic) speeds beyond knot F ($\sim 700$ pc) \citep[see, e.g.,][]{Meyer2017}. Taking $\Gamma_{\rm VLBI}/\Gamma_{\rm kpc} = 1$ in Eq.~(\ref{eq:bkpc}), upper limits of $B_{\rm kpc}$ are $\sim 200\, \mu$G (for a toroidal field), and $\sim 0.02\, \mu$G (for a poloidal field). An independent estimate of the magnetic field in the M87 jet is obtained from an analysis of radio and optical images which suggest values in the range $20-40$ $\mu$G \citep[][]{Heinz1997}. However, recent constraints coming from the upper limit on $\gamma$-ray favour values nearer to the equipartition value of $B_{\rm eq} \sim 300 \, \mu$G at knot A \citep[][]{Stawarz2005}. In summary, we can consider that the magnetic field strengh at 1~kpc for the M87 core is in the range 30-300 $\mu$G (suggesting a range of electron density between n$_e$ $\sim$ 0.2 [cm$^{-3}$] for B=30 $\mu$G and 0.02 [cm$^{-3}$] for B=300 $\mu$G). Now, the comoving magnetic energy density associated to this range is $U_B = B^2/8\pi \approx 3.2 \times 10^{-11} - 3.2 \times 10^{-9}$ erg/cm$^{3}$. These values can be now compared with the total energy density associated with the bulk of the energy flux carried by the matter of the M87 jet, $P_{\rm j}$, 
\begin{equation}
    U_{\rm P} \approx \frac{P_{\rm j}}{\pi R_{\rm kpc}^2 \Gamma_{\rm kpc}^2 c}.
    \label{eq:up}
\end{equation}

For the kinetic power of the M87 jet, we use the estimate of the total kinetic luminosity of the jet based on the examination of the energy stored in the inner radio halo \citep[$\sim$ 5 kpc in extent, as derived by][]{Reynolds1996}. This estimate, $P_{\rm j,\, min} = 1.0 \times 10^{43}$ erg/s, is a reliable lower bound of the total kinetic energy transported by the jet since it neglects the radiated energy, the work exerted on the surrounding interstellar medium to expand the radio halo, and the contribution of the thermal particle energy (which can dominate the kinetic energy flux beyond HST-1 due to entrainment). Using the minimum estimated jet power in Eq.~(\ref{eq:up}) gives $U_{\rm P, \, min} \approx 1.0 \times 10^{-9} - 3.0 \times 10^{-9}$ erg/cm$^3$, where we have used $R_{\rm kpc} = 22, 62$ pc for knots D and A \citep{Owen2000}, and $\Gamma_{\rm kpc} = 5, 1$ \citep[knots D and A, respectively;][]{Meyer2017}. Our results give $U_B \lesssim U_{\rm P, \, min}$, hence supporting the idea that the M87 jet is, most likely, particle dominated and justifying an interpretation of the helical jet structure based on (purely hydrodynamic) K-H modes \citep{Lobanov2003, Hardee2011}. 

\subsection{A helical magnetic field maintained by K-H instabilities}

The next issue is to explain the helical configuration of the magnetic field in the kiloparsec scale jet of M87 with a dominant poloidal component as seen in Fig.~\ref{Fig2}. Current jet formation mechanisms rely on the existence of intense poloidal fields in the vicinity of the central black hole \citep[see, e.g.,][]{Komissarov2012}. These fields can be generated in the inner parts of the accretion disk from the azimuthal magnetic field as buoyant magnetic loops with small length scales (of the order of the disk thickness) emerging out of the disk. Additionally, large scale poloidal magnetic fields can be generated by accumulation of magnetic field from the interstellar medium. However, in the standard acceleration mechanism, there is a net conversion of poloidal field into toroidal one within the acceleration region. Moreover, the upper bound of 0.2 G for the magnetic field at VLBI scales and the magnetic flux conservation leads to stringent limits on the poloidal magnetic field at kpc scales incompatible with this component as being dominant and responsible of the synchrotron emission. In alternative scenarios \citep{Heinz2000}, jets are accelerated by tangled magnetic fields. In these models, processes like turbulence and small scale instabilities drive a continuous transfer of energy from the slowly decaying transverse (toroidal, radial) component of the magnetic field to the rapidly decaying longitudinal component leading to magnetic fields with toroidal and poloidal components of similar strengths but fully randomized.

It is, nevertheless, plausible that the observed magnetic field is enhanced by the pressure maxima caused by the K-H instability modes. Within this maxima, the magnetic field lines would also be compressed, the emission enhanced, and therefore, also its polarized emission. Altogether, the ordered magnetic field structure would be caused by the order imposed by the instability modes pressure maxima, in agreement with the minor dynamical role of the magnetic field at the observed scales. In addition, the fact that these are surface modes \citep[][]{Lobanov2003,Hardee2011} would facilitate the compression of the poloidal field generated by shear at the jet-ambient transition region.

Recent work by \cite{Beuchert2018} and \cite{Schulz2020} have interpreted both kinematics and polarization structure of the jet in the radio-galaxy 3C111 as caused by the different axial velocity of the jet plasma on axis and at the jet boundaries. This effect has also been reported by a number of numerical simulations both in non-relativistic \citep[][]{Matthews1990,Gaibler2009,Huarte2011,Hardcastle2014} and relativistic jets \citep[][]{RocaSogorb2009,Marti2016}. Furthermore, \cite{Laing2014} also showed evidence of jet deceleration at the boundaries prior to the axis in classical Fanaroff and Riley Class I (FRI) jets, which also results in shear between those regions. Actually, the fitted longitudinal fields at the jet boundaries is still non-negligible along the decelerating region for the studied FRI jets \citep[see Figs. 19 and 20 in ][]{Laing2014}, but this seems to be strongly dependent on the dynamics of the individual sources. Therefore, we can claim that in the case of M87 the evolution of surface modes would take place in a region where the magnetic field has a predominant poloidal component. Compression produced by the instability modes would then brighten up the affected regions.

\subsection{Concluding remarks}

The jet of M87 has the remarkable characteristic of showing several bright knots along its path, from the pc to kpc scales. Since early studies, these knots have been proposed to actually trace filamentary structures, most probably a consequence of K-H instabilities. This morphology of a jet characterized by bright knots up to kpc scale is very clear in M87, mainly because of its brightness and its proximity. But, similar characteristics have been proposed in several other jets. Remarkably in 3C273, a filamentary structure due to K-H instabilities have also been proposed \citep[][]{Lobanov2001, Perucho2006}. The unprecedented high angular resolution and high sensitivity images obtained in this work allowed to clearly resolve, for the first time, the morphology of a large part of the M87 jet into a double-helix structure, in well agreement with expectations from K-H instabilities. Thus, we confirm that the knots in this object are simply the locations where the two filaments cross each other. Not only that, as our analysis of M87 suggests, it is the presence of this double-helix structure what actually allows to maintain a helical magnetic field to larger scales. Therefore, we speculate that other objects driving bright jets to kpc scales, when observed with enough high angular resolution and sensitivity, will also show a similar double helix morphology. With the current instrumentation, only the brightest and close-by objects might be suitable targets for such a study. However, the future high sensitive radio telescopes, i.e. SKA or ngVLA, will allow to explore the connection between instabilities and magnetic fields in AGN jets.



\begin{acknowledgments}
ACKNOWLEDGMENTS\\
A.P. acknowledges the program "Programa de Investigadoras e Investigadores por M\'exico" (CONACyT). JLG and AF acknowledge financial support from the Spanish Ministerio de Econom\'{\i}a y Competitividad (grants AYA2016-80889-P, PID2019-108995GB-C21), the Consejer\'{\i}a de Econom\'{\i}a, Conocimiento, Empresas y Universidad of the Junta de Andaluc\'{\i}a (grant P18-FR-1769), the Consejo Superior de Investigaciones Cient\'{\i}ficas (grant 2019AEP112), and the State Agency for Research of the Spanish MCIU through the Center of Excellence Severo Ochoa award for the Instituto de Astrof\'{\i}sica de Andaluc\'{\i}a (SEV-2017-0709).
JMM and MP acknowledge support from the Spanish \emph{Ministerio de Ciencia} through grant PID2019-107427GB-C33, and from the \emph{Generalitat Valenciana} through grant PROMETEU/2019/071. JMM acknowledges additional support from the Spanish \emph{Ministerio de Econom\'{\i}a y Competitividad} through grant PGC2018-095984-B-100. MP acknowledges additional support from the Spanish \emph{Ministerio de Ciencia} through grant PID2019-105510GB-C31.
\end{acknowledgments}

\bibliography{Biblio.bib}{}
\bibliographystyle{aasjournal}

\end{document}